\documentclass[preprint,10pt]{b_test}
\usepackage{amssymb}
\usepackage{amsmath}
\usepackage{amsthm}
\usepackage{graphicx}
\usepackage{url}
\usepackage{hyperref}
\usepackage{hypcap}
\usepackage{natbib}

\newtheorem{mydef}{Definition}
\newtheorem{prop}{Proposition}
\newtheorem{pr}{Property}
\newtheorem{as}{Assumption}
\newtheorem{thm}{Theorem}

\newtheorem{co}[thm]{Corollary}
\newdefinition{rmk}{Remark}
\newproof{pf}{Proof}
\newproof{pot}{Proof of Theorem \ref{thm2}}
\journal{Systems and Control Letters}
\usepackage{natbib}
\usepackage{tikz}
\usetikzlibrary{arrows,shapes,snakes,backgrounds}
\usepackage{pgfpages}
\tikzstyle{every picture}=[>=stealth,thick] % Fine piler og feite linjer er bra i presentasjonar.
\usepackage{varioref}
\usepackage{geometry}
\usepackage{graphicx}

\begin{document}

\begin{frontmatter}
\title{Measuring and Analysing Marginal Systemic Risk Contribution using $CoVaR$: A Copula Approach}
\author{Brice Hakwa}
\ead{hakwa@uni-wuppertal.de}
\address{Fachbereich C - Mathematik - Stochastik, Bergische Universit\"at Wuppertal}
\author{Manfred J\"ager-Ambro\.{z}ewicz }
\ead{Manfred.Jaeger-Ambrozewicz@HTW-Berlin.de}
\address{Hochschule f\"ur Technik und Wirtschaft Berlin - University of Applied Sciences HTW Berlin}
\author{Barbara R\"udiger}
\ead{ruediger@uni-wuppertal.de}
\address{Fachbereich C - Mathematik - Stochastik, Bergische Universit\"at Wuppertal} 
\begin{abstract}
This paper is devoted to the  quantification and  analysis  of the marginal risk contribution of a given single financial institution $i$ to the risk of a financial system $s$. Our work expands on the  $CoVaR$ concept  proposed  by Adrian and Brunnermeier \cite{covar} as a tool for the measurement of  marginal systemic risk contribution. We first give a mathematical definition of $CoVaR_{\alpha}^{s|L^i=l}$. Our definition improves the $CoVaR$ concept by expressing $CoVaR_{\alpha}^{s|L^i=l}$  as a function of a state $l$ and of a given probability level $\alpha$ relative to   $i$ and $s$ respectively. Based on copula theory we connect $CoVaR_{\alpha}^{s|L^i=l}$ to the partial derivatives of Copula  through their  probabilistic  interpretation (Conditional Probability). Using this we provide a closed formula for  the calculation of $CoVaR_{\alpha}^{s|L^i=l}$ for a large class of (marginal) distributions and  dependence structures (linear and non-linear). Our formula allows a better analysis of systemic risk using $CoVaR$ in the sense that it allows us to define $CoVaR_{\alpha}^{s|L^i=l}$ depending on the marginal distributions of the  losses $L^i$ and $L^s$ of $i$ and $s$ respectively on the one hand and the copula of $L^i$ and $L^s$ on the other hand. We discuss the implications of this in the context of the quantification and analysis of systemic risk contributions. %some mathematical This makes possible the
We will, for example, highlight  some of the effects of the marginal distribution $F_s$ of $L^s$, the dependence parameter $\rho$, and the condition $C\left(L^i\right)$ on $CoVaR_{\alpha}^{s|C\left(L^i\right)}$.  

 \end{abstract}
 \begin{keyword} $VaR$\sep $CoVaR$\sep  Systemic risk\sep Copula\sep Conditional Probability   \\[0.2cm] 
\end{keyword}
 
\end{frontmatter}

%%%%%%%%%%%%%%%%%%%%%%%%%%%%%%%%%%%%%%%%%%%%%%%%%%%%%%%%%%%%%%%%%

%%%%%%%%%%%%%%%%%%%%%%%%%%%%%%%%%%%%%%%%%%%%%%%%%%%%%%%%%%%%%%%%%

 \section{Introduction}
%The failure of certain finnancial institutions such as Lehman-Brother during the recent crisis highlighted the significant of the adverse impact that a failure of a single firm can have on the
%financial system as a whole.
With the last crisis it became clear that the failure of certain financial institutions (the so called system relevant financial institutions) can produce an adverse impact on whole financial system.  The inability of standard risk-measurement tools like   Value-at-Risk ($VaR$)  to capture this systemic nature of risk (since their focus is on an institution in isolation: micro risk management)  poses a new risk-management challenge  to the financial regulators  and  academics. We can summarise this into two questions:
\begin{enumerate}
	\item How to identify   System-relevant Financial Institutions ?
	\item How to quantify the marginal risk contribution of one single financial institute to  the system ?
\end{enumerate}
As an academic response to this problems, Adrian and Brunnermeier  proposed $CoVaR$  (\cite{covar}) as a model to  analyse the marginal adverse financial effect of a distressed  single financial institution  on  the financial system. %the contributions that individual banks make to systemic risk by analysing the state of the financial system under the condition of a single institution being in distress.  
They defined the risk measure  $CoVaR$ as the Value at Risk ($VaR$) of the financial system conditional to the state of the loss of a single institution  and  quantify the institution's marginal risk contribution (how much an institution adds to the risk of the system) by the measure $\Delta CoVaR$. This is defined as the difference between $CoVaR$ conditional to the institution under distress and the $CoVaR$ conditional to the institution  in a normal state.

 %$CoVaR$ quantifies how much an individual institution $i$ contributes to the systemic risk $s$. Its computation is based  on

Thus the implementation of $CoVaR$  involves  variables characterising a single financial institution $i$ (e.g. $L^i$) and the financial system $s$ (e.g. $L^s$) respectively and variables characterising the interdependency structure within the financial system and  between single financial institutions  and the financial system $s$.  This macro-dimension of $CoVaR$  allows the integration of the dependence structure of $i$ and $s$ in the risk-measurement contrary to the standard risk measures ("micro-risk measure" e.g. VaR) where only variables characterising the financial institution alone are considered. The $CoVaR$ concept can be thus used by regulatory institutions  as a macro-prudential tool (or as a basis for the development of other tools) to identify systemically relevant financial institutions and to set the adequate capital requirements. % on the basis of certain firm-specific characteristics like its loss, its size and its correlation with the whole system. %However there are some gaps in its ability to describe and integrate  distressed markets.

But its calculation represents an open problem.

But its computation represents an open problem. Although some approaches have been proposed, \citet{covar} proposed for example an estimation method based on  ''linear quantile regression'', \citet {lehar1} adopted a simulation based approach, \citet{manfred} developed a closed formula for the special case that the joint distribution of financial system characteristic variable is of the Gaussian type. In all these  approaches there are some difficulties   to flexibly model the  stochastic behaviors of financial institution's specific variables and their dependence structure (interconnection) within a financial system, since only linear dependence are considered. \\  
Our aim is thus to provide a more flexible framework for the implementation of the $CoVaR$ concept which allows the integration  of stylised features of marginal losses  as  skewness, fat tails and interdependence properties like linear, non-linear and positive or negative tail dependence. To do this we first propose an improved definition of $CoVaR$ which makes it mathematically tractable (see def.~\ref{def}), and based on copula theory we propose a general analytical formula for $CoVaR$ (see Theorem~\ref{main}). We use our formula to make some theoretical  analyses and computations related to  $CoVaR$. \\ %We do this by proposing a copulas based framework    for   $CoVaR$.% and by   defining  a new systemic risk measure namely the $TCoVaR$ (respectively $\Delta TCoVaR$) %and derive some related risk contribution measures.
We conclude this article by applying our formula to compute the $CoVaR_{\alpha}^{s|L^i=l}$in the Gaussian copula (Section~\ref{ex1}), t-copula (Section~\ref{ex2}), and Gumbel copula setting (Section~\ref{ex3}) respectively. We discuss the results of our computation and draw from this some interesting conclusions. %(for example the relation between $VaR$, $CoVaR_{\alpha}^{s|L^i=l}$, $\rho$ and $\Delta CoVaR$). 
We also give a general formula for  $CoVaR_{\alpha}^{s|L^i=l}$in the Archimedian copula setting (Section~\ref{ex4}). %i.e. We show that in the Gaussian framework the $CoVaR$ and the $\Delta CoVaR$ may be expressed as a linear function of the value at risk ($VaR^i$) of the corresponding single financial institution $i$.

\subsection{\textbf{Definition of $CoVaR$ and $\Delta CoVaR$}}
\vspace{0.2cm}
We recall here the definition of the value at risk ($VaR$) in order to define the $CoVaR^{s|i}$ as a conditional $VaR$ following Adrian and Brunnermeier (\cite{covar}).   
\begin{mydef}[Value at Risk] Given some confidence level $\alpha \in \left(0,1\right)$ the $VaR$ of 
a portfolio at the confidence level $\alpha$ is given by the smallest number $l$ such that the probability that the loss $L$  exceeds $l$  is no larger than $\left(1-\alpha\right)$. Formally 
\begin{align*}
	VaR_{\alpha}&:=\inf\left\{l \in \mathbb{R}: Pr\left(L>l\right)\leq 1 - \alpha\right\}& \\
	&=\inf\left\{l\in\mathbb{R}: Pr\left(L\leq l\right)\geq \alpha\right\}.& 
\end{align*} 
\end{mydef}
 In order to give a  probabilistic interpretation of $VaR_\alpha$,  we will employ the notation of quantiles as provided in the following definition (cf. \cite{qrm} def.~2.12).
 \begin{mydef} [Generalised inverse and quantile function ]
\hspace{0.5cm}
\begin{enumerate}
	\item Given some increasing function $T : \mathbb{R}\rightarrow \mathbb{R}$, the generalised inverse of $T$ is
defined by $T(y):=\inf\left\{x \in \mathbb{R}: T\left(x\right)\geq y\right\}$.  
\item Given some distribution function $F$, the generalised inverse $F^{\leftarrow}$ is called the quantile function of $F$. For $\alpha \in \left(0, 1\right)$ we have
\begin{align*}
	q_{\alpha}\left(F\right)=F^{\leftarrow}\left(\alpha\right):=\inf\left\{x \in \mathbb{R}: F\left(x\right)\geq \alpha\right\}.
\end{align*}
\end{enumerate}
\end{mydef}
Note that, if $F$ is continuous and strictly increasing, we simply have 
\begin{align}
	q_{\alpha}\left(F\right) = F^{-1}\left(\alpha\right),
\end{align}
where  $F^{-1}$ is the (ordinary) inverse of $F$.  
Thus suppose that the distribution $F$ of the loss $L$ is continuous and strictly increasing. It follows  
\begin{align}
	\normalfont{VaR}_{\alpha}=F^{-1}\left(\alpha\right).
\end{align}
We note that typical values taken  for $\alpha$ are  0.99 or 0.995. \\
\begin{as}
\label{as}
Henceforth we consider  only random variables which have  strictly positive density function. Also in case we consider a bivariate joint distribution $H\left(x,y\right)$ we assume that it has a density and its marginal distributions have strictly positive densities. 
\end{as}
So due to this assumption  all considered   distribution functions $F$ are continuous and strictly increasing. Such an  $F$ is thus  invertible and  $F^{-1}$ denotes the unique inverse of  $F$.  
Let $L^i$ be the loss  of the financial institution   $i$ and $L^s$ the loss of the system $s$ without the institution $i$.  
At least since the financial crisis it is clear that the dependency between the system and the institution $i$ must be analysed more seriously. A step towards such an analysis is done by  explicitly defining $CoVaR$. 
Adrian and Brunnermeier  denote by $CoVaR_{\alpha}^{s|C\left(L^i\right)}$ the value of an institution $s$ (or a financial system)
conditional on some event $C\left(L^i\right)$  depending on the loss  $L^i$ of an institution $i$. Thus $CoVaR_{\alpha}^{s|C\left(L^i\right)}$ can be implicitly defined as the $\alpha-quantile$ of the conditional probability of the system's  loss.
\begin{align}
	Pr\left(L^s\leq CoVaR_{\alpha}^{s|C\left(L^i\right)}|C\left(L^i\right)\right)=\alpha.
	\label{eq:3}
\end{align}
 They analysed in their work  \cite{covar}  the case that the condition $C\left(L^i\right)$ refers to the loss $L^i$ of institution $i$ being exactly at its value at risk or more generally being exactly at some specific value $l$. We have in this case in the context of \eqref{eq:3} the following expression,
 \begin{align}
	Pr\left(L^s\leq CoVaR_{\alpha}^{s|L^i=l}| L^i=l\right)=\alpha.
 	\label{eq:4}
\end{align}
Due to assumption~\ref{as} 
 \begin{align*}
	Pr\left(L^i = l\right) = 0,   \text{ for any } l\in \mathbb{R}.
\end{align*}
However we can define in the context of assumption~\ref{as}, a conditional probability of the form: $Pr\left(L^s \leq h |L^i=l\right)$ for fixed $l$  as a function of $h$ as follows [cf. \cite{leo} p.~72) or\cite{feller} p.~71 ].  
\begin{align}
\label{eq:5}
	Pr\left( L^s \leq h |L^i=l \right) & =: R_l \left(h\right)& \nonumber \\
  &= \int_{- \infty}^{h}  \frac{f\left(l,y\right) }{f_{i}\left(l\right) } dy.   &
\end{align}
Where $f_{i}\left(x\right)= \int_{-\infty}^{+\infty}f\left(x,y\right)dy$ is the marginal density of $L^i$. \\
Note that $\eqref{eq:5}$ is defined only when $f_{i}\left(l\right) \neq 0$; however, if $\mathcal{S} = \left\{\left(l,y\right): f_i\left(l\right)\neq 0\right\}$, then $Pr\left(\left(L^i,L^s\right)\in \mathcal{S}\right)=0$.
 \begin{rmk}
Due to assumption~\ref{as} we have that,
\begin{itemize}
	\item  the functions  $R_l$ is  well defined. (since  $f_i\left(l\right) > 0, \forall \ l\in \mathbb{R} $),% $\eqref{eq:5}$ is defined only when $f_{i}\left(l\right) \neq 0$
	\item   $R_l\left(h\right)$ 	is strictly  increasing and continuous.
\end{itemize}
\end{rmk}
 As $R_l\left( h\right)$ is strictly  increasing, it follows that its is invertible. Based on this we provide a alternative  definition for $CoVaR_{\alpha}^{s|L^i=l}$ which is more tractable from a mathematical point of view than that proposed by Adrian and Brunnermeier. 
\begin{mydef}
\label{def}
Assume that $L^i$ and $L^s$ have density which satisfy assumption~\ref{as} .% strictly positive continuous joint density function $f$. 
Then for a given $\alpha \in \left(0,1\right)$ and for a fixed $l$, $	CoVaR_{\alpha}^{s|L^i=l}$ is defined as:
\begin{align}
	CoVaR_{\alpha}^{s|L^i=l}&:= \inf \left\{h \in \mathbb{R}: \ Pr\left(L^s > h | L^i =l \right)\leq 1-\alpha\right\}& \nonumber \\
	                          &:= \inf \left\{h \in \mathbb{R}: \ Pr\left(L^s \leq h | L^i =l \right)\geq \alpha\right\}& \nonumber \\
	                          &=R_l^{-1}\left(\alpha\right).&
\end{align}
\end{mydef}
 \begin{mydef}[$\Delta CoVaR^{s|i}$]
 \label{d4}
Adrian and Brunnermeier  denote by $\Delta CoVaR_{\alpha}^{s|i}$ the difference between $CoVaR_{\alpha}^{s|C\left(L^i\right)}$ condition on the institution $i$ being under distress and the $CoVaR_{\alpha}^{s|C\left(L^i\right)}$  condition on the institution having mean loss.
\begin{align}
\Delta CoVaR^{s|i}_{\alpha}= CoVaR^{s|L^i=VaR^i_{\alpha}}_{\alpha} - CoVaR^{s|L^i = E\left(L^i\right)}_{\alpha}.
\end{align}
$\Delta CoVaR^{s|i}$ is used as measure  to quantify  the marginal risk contribution of a single institution $i$ to the  risk of the system.
\end{mydef}
We will find in the next a closed analytical formula in terms of copula in the context of definition \ref{def} (see Theorem~\ref{main}).

%%%%%%%%%%%%%%%%%%%%%%%%%%%%%%%%%%%%%%%%%%%%%%%%%%%%%%%%%%%%%%%%%

%%%%%%%%%%%%%%%%%%%%%%%%%%%%%%%%%%%%%%%%%%%%%%%%%%%%%%%%%%%%%%%%%

 \section{A Brief Introduction to Copulas }
%%%%%%%%%%%%%%%%%%%%%%%%%%%%%%%%%%%%%%%%%%%%%%%%%%%%%%%%%%%%%%%%%%%%%%%%%%%%%
In this section we introduce the notion of copula and give some basic  definitions and  important properties needed later.  Our focus is on properties that will be helpful when connecting copulas to conditional probabilities and analyzing $CoVaR_{\alpha}^{s|L^i=l}$ and $\Delta CoVaR^{s|i}$ (for detailed analysis of copulas, we refer the reader to e.g. \cite{joe}, \cite{qrm},  \cite{nelsen} or \cite{ronc2} and the references therein).
\subsection{Preliminary}
\vspace{0.2cm}
In order to introduce  the concept of a copula, we recall some important remarks upon which it is built.
\begin{rmk}[cf. \cite{qrm} proposition.~5.2]
\label{prop:2}
\hspace{0.2cm}
\begin{enumerate}
	\item \textbf{Quantile transformation.}  If $U \sim U\left(0,\ 1\right)$ is standard uniform distributed, then 
\begin{align*}
	Pr\left(F^{-1}\left(U\right)\leq x\right)=F\left(x\right).
\end{align*}
	\item \textbf{Probability transformation.} Assume $F$ is a distribution function such that its inverse function $F^{-1}$ is well defined.  Let $X$ be a random variable with distribution function $F$, then $F(X)$ has a uniform standard distribution  
\begin{align*}
	F(X)\sim U\left(0, 1\right).
\end{align*}
\end{enumerate}
\end{rmk}
 %%%%%%%%%%%%%%%%%%%%%%%%%%%%%%%%%%%%%%%%%%%%%%%%%%%%%%%%%%%%%%%%%%%%%%%%%%%%
\subsection{\textbf{Definition and basic properties of Copula}} 
%%%%%%%%%%%%%%%%%%%%%%%%%%%%%%%%%%%%%%%%%%%%%%%%%%%%%%%%%%%%%%%%%%%%%%%%%%%%%%
\vspace{0.2cm}
 
\begin{mydef}[2-dimensional copula (cf. \cite{nelsen} def.~2.2.2)]
A 2-dimensional copula is a (distribution) function $C: \left[0, 1\right]^2\rightarrow \left[0,1\right]$ with the following satisfying:
\begin{itemize}
	\item Boundary conditions:
	   \begin{enumerate}
	\item[1)] For every $u \in \left[0, 1\right] : C\left(0,u\right) = C\left(u,0\right)=0.$
		\item[2)] For every $u \in \left[0, 1\right] : C\left(1,u\right) = u \ and  \ C\left(u,1\right)=u.$
		\end{enumerate}
		\item Monotonicity condition:
\begin{enumerate}
\item[3)] 	 For every $\left(u_1,u_2 \right),\left(v_1,v_2 \right) \in \left[0, 1\right] \times \left[0, 1\right] with \ u_1\leq u_2 \  and \ v_1 \leq v_2 $ we have
 \begin{align*}
		C\left(u_2,v_2\right)-C\left(u_2,v_1\right)-C\left(u_1,v_2\right)+C\left(u_1,v_1\right)\geq 0.
	\end{align*}
	\end{enumerate}
\end{itemize}	
\end{mydef}
Conditions (1) and (3) implies that the so defined 2-copula C is a bivariate joint distribution function (cf. \cite{nelsen} def.~2.3.2) and condition (2) implies that the copula C has standard uniform margins. 
We present now some important basic properties of copulas which we will use below (cf. \cite{nelsen} chap.~2).
   All this is summarised
in the following theorem.
\begin{thm}[cf. \cite{nelsen} Thm.~2.2.7] Let C be a copula. For any $v \in \left[0, 1\right]$, the partial derivative $\partial C\left(u, v\right)/\partial u$ exists for almost all  $u$, and for such $v$ and $u$
\begin{align*}
	0\leq \frac{\partial C\left(u,v\right)}{\partial u}\leq 1.
\end{align*}
Similarly, for any $u \ \in \left[0, 1\right]$, the partial derivative $\partial C\left(u, v\right)/\partial v$ exists for almost all $v$, and for such u and v
\begin{align*}
	0\leq \frac{\partial C\left(u,v\right)}{\partial v}\leq 1.
\end{align*}
Furthermore, the functions $ u \mapsto \partial C\left(u, v\right)/\partial v$   and $ v \mapsto \partial C\left(u, v\right)/\partial u$ are
defined and nondecreasing  everywhere on $\left[0, 1\right]$.
\end{thm} 
The following theorem  makes the copula theory attractive as tool for stochastic modeling because it links joint distributions to one-dimensional marginal distributions.   
\begin{thm}[Sklar's theorem, cf.   \cite{nelsen} Thm.~2.3.3]Let $H$ be a joint distribution function with marginal distribution functions $F \ and \ G$. Then there exists a copula $C$ such that for all $x,y \in \mathbb{R}\cup \left\{-\infty\right\} \cup \left\{+\infty\right\}$
	\begin{align}
	 \label{eq:10}
		H\left(x,y\right)=C\left[F\left(x\right),G\left(y\right)\right].		             
	\end{align}  
If $F$ and $G$ have density, then $C$ is unique. Conversely, if $C$ is a copula and $F$ and $G$ are distribution functions, then the function $H$ defined by \eqref{eq:10} is a joint distribution function with margins $F$ and $G$.
\end{thm}
This theorem is very important because it asserts that, using copula function, it is possible  to  represent each bivariate distribution function as a function of univariate distribution function.  
Thus, we can use the copula to extract the dependence structure among the components $X$ and $Y$ of the vector $\left(X, Y \right)$, independently of the marginal distribution $F$ and $G$. This allows us to   model the dependence structure and marginals  separately.
\begin{rmk}
Assume   $\left(X, Y\right)$ is a bivariate  random variables  with  copula $C$ and joint distribution $H$ satisfying assumption~\ref{as}, with marginals distribution function $F$ and $G$.   Then the transformed randoms variables $ U=F\left(X\right)$ and $V=F\left(Y\right)$ have standard uniform distribution  and $C\left(U, V\right)$ is the joint distribution of $\left(U, V\right)$. In fact
\begin{align*}
	C \left(u, v\right) = C \left(Pr\left(U\leq u\right), Pr \left(V\leq v \right)\right).
\end{align*}
 \end{rmk}
 \begin{co}[cf. e.g. \cite{nelsen} co.~2.3.7]
Let $H$ denote a bivariate  distribution function with   margins $F$ and $G$ satisfying assumption~\ref{as} .  
Then there exist a unique copula $C$ such that  for all  $(u,v)\  \in \left[0, 1\right]^2$ it holds:
\begin{align*}
	C\left(u,v\right)=H\left(F^{-1}\left(u\right),G^{-1}\left(v\right)\right).
\end{align*}
\end{co}

%%%%%%%%%%%%%%%%%%%%%%%%%%%%%%%%%%%%%%%%%%%%%%%%%%%%%%%%%%%%%%%%%

%%%%%%%%%%%%%%%%%%%%%%%%%%%%%%%%%%%%%%%%%%%%%%%%%%%%%%%%%%%%%%%%%

 \section{Computing and Analysing  systemic Risk Contribution with $CoVaR_{\alpha}^{s|L^i=l}$:  A Copula Approach} 
\label{sec:3} 
In this section we provide a copula based framework for the calculation and the theoretical analysis  of $CoVaR_{\alpha}^{s|L^i=l}$ as tool for the measurement of systemic risk contribution.  
To do this  we will relate the notion of conditional probability to copulas and rewrite the implicit definition of    $CoVaR^{s|L^i=l}$   in terms of copula.  
Based on this  we will  derive some  useful  results.  Specifically,  we  will obtain a closed  formula which will provide a general  framework for the flexible   calculation and analysis of  $CoVaR^{s|L^i=l}$   in many stochastic settings. Based on this formula we will highlight some important properties of $CoVaR_{\alpha}^{s|L^i=l}$ and $\Delta CoVaR^{s|i}$. 
\subsection{\textbf{Computation of  $CoVaR_{\alpha}^{s|L^i=l}$ using Copula  }}
\vspace{0.2cm}
We propose here in the following theorem  a general framework  for  computing  $CoVaR_{\alpha}^{s|L^i=l}$ analytically. Our approach is based on  the copula representation of conditional probability. 
\begin{thm}
 \label{main}
Let $L^i$ and $L^s$ be two random variables representing the loss of the system $i$ and institution $s$ with  marginal distribution functions $F_s$ and $F_i$ respectively. Let $H$  be the joint distribution of  $L^i$ and $L^s$  with the corresponding  bivariate copula $C$, i.e.
\begin{align*}
	H\left(x,y\right)=C\left(  F_i\left(x\right),F_s\left(y\right)\right).
\end{align*}
Let us assume assumption~\ref{as}  and  
\begin{align*}
	g\left(v, u\right):=\frac{\partial C\left(u, v\right)}{\partial u}
\end{align*}
is invertible  with respect to the  parameter $v$. Then for all $l\in \mathbb{R}$ $CoVaR_{\alpha}^{ s|L^i=l}$ at level $\alpha,\   0 < \alpha < 1 $ is given by  
\begin{align}
	CoVaR_{\alpha}^{ s|L^i=l}\left(\alpha\right)=F_s^{-1}\left(g^{-1}\left(\alpha,F_i\left(l\right)\right)\right)  \  \forall \ \alpha \in \left[0, 1\right].
	\label{eq:18}
\end{align}
\end{thm}
\begin{pf}  
Recall that the implicit definition of $CoVaR_{\alpha}^{s|L^i=l}$ is given by:
\begin{align*}
	&Pr\left( L^s\leq CoVaR_{\alpha}^{s|L^i=l}|L^i=l\right)=\alpha &\\
	\Leftrightarrow  &Pr\left(F_s\left(L^s\right)\leq F_s\left(CoVaR_{\alpha}^{s|L^i=l}\right)|F_i\left(L^i\right)=F_i\left(l\right)\right)=\alpha.&
	\end{align*}
Let $V=F_s\left(L^s\right),\ \ U=F_i\left(L^i\right),\ \ v=F_s\left(CoVaR_{\alpha}^{s|L^i=l}\right)$  and $u=F_i\left(l\right)$ i.e. 
\begin{align*}
	Pr\left( L^s\leq CoVaR_{\alpha}^{s|L^i=l}|L^i=l\right)&=Pr\left(F_s\left(L^s\right)\leq F_s\left(CoVaR_{\alpha}^{s|L^i=l}\right)|F_i\left(L^i\right)=F_i\left(l\right)\right)&\\
	                    &=Pr\left(V\leq v|U=u\right).&\\
\end{align*}
 Due to assumption~\ref{as} it follows from remark \ref{prop:2} that $V$ and $U$ are  standard uniform distributed. 
In this case we can refer to  (\cite{leo} eq.~(4.4)) and  (\cite{ronc2} p.~263)) and compute the conditional probability $Pr\left(V\leq v|U=u\right)$, as follows:        
	\begin{align*}            %   &=\frac{Pr\left(V\leq v, U=u\right)}{Pr\left(U=u\right)}&\\
	                 Pr\left(V\leq v|U=u\right)&=\lim_{\Delta u \to 0^+}\frac{Pr\left(V\leq v, u\leq U \leq u +\Delta u \right)}{Pr\left(u\leq U \leq u +\Delta u \right)}&\nonumber\\                    
	                 &=\lim_{\Delta u \to 0^+}\frac{Pr\left(U \leq u+ \Delta u,V  \leq v\right) -  Pr\left(U\leq  u ,V  \leq v\right)}{Pr\left( U \leq u +\Delta u  \right)-Pr\left(U \leq u  \right)}&\nonumber\\ 
	                 &=\lim_{\Delta u \to 0^+}\frac{C\left(u+ \Delta u,v\right) -  C\left(u ,v\right)}{\Delta u }&\nonumber\\ 
	                  &=\frac{\partial C\left(u,v\right)}{\partial u }=g\left(u, v\right).&     %\label{dev1}                                  
\end{align*}
Now we are able to derive the explicit expressions of $CoVaR_{\alpha}^{s|L^i=l}$ provided that, 
 the function $g$ is invertible with respect to the ''non-conditioning'' variable $v$. In this case we can write $v$ as a function of    $\alpha$, $u$ as follow 
\begin{align*}
	v=g^{-1}\left(\alpha,u\right).
\end{align*}
Using  $v=F_s\left(CoVaR_{\alpha}^{s|L^i=l}\right)$  and $u=F_i\left(l\right)$ we obtain

\begin{align*}
F_s\left(CoVaR_{\alpha}^{s|L^i=l}\right)=g^{-1}\left(\alpha, F_i\left(l\right)\right).
\end{align*}
Thus
\begin{align*}
	&CoVaR_{\alpha}^{s|L^i=l}=F_s^{-1}\left(g^{-1}\left(\alpha, F_i\left(l\right)\right)\right).& \blacksquare 
\end{align*}
\end{pf}
In practice the conditional level $l$ for the financial institution  $i$ is implicitly defined by a given confidence level $\beta$ such that 
\begin{align}
 \label{eq:22}
	 l=F_i^{-1}\left(\beta\right), %and $F\left(l\right)=\¦\alpha,
\end{align}
$\beta$ is specified by the regulatory institution. It represents the probability with which the financial institution $i$ remains solvent over a given period of time horizon. %By taking this information in consideration and suppose that $\alpha_i$ be the confidence level required by the regulatory institution for the financial institution $i$.  
Base on this information we can express $CoVaR_{\alpha}^{s|L^i=l}$ as follow:
 \begin{align}
	CoVaR_{\alpha}^{s|L^i=l}=F_s^{-1}\left(g^{-1}\left(\alpha, \beta\right)\right).\label{eq:23}
\end{align}
We remark that for a given marginal distribution of the system's losses $F_s$ the above  expression of  $CoVaR_{\alpha}^{s|L^i=l}$ has only as input parameter  $\alpha$  and $\beta$. This motivates the following definition.
\begin{mydef}
\label{d6}
\begin{align*}
			CoVaR_{\alpha}^{\beta} := CoVaR_{\alpha}^{s|L^i=l}
\end{align*}
\end{mydef}
\begin{rmk}
Equation~\eqref{eq:23} is very important because it asserts that in the practice $CoVaR_{\alpha}^{s|L^i=l}$ (or $CoVaR_{\alpha}^{\beta}$ ) contrary to standard risk-measurement tools like   Value-at-Risk ($VaR$) does not depend of the marginal distribution $F_i$ but depends only on the marginal distribution of the system's losses $F_s$  and the copula between the financial institution $i$ and the financial system $s$. 
\end{rmk}
 \begin{rmk}
\label{r2}
We can see from equation \eqref{eq:18} that $CoVaR_{\alpha}^{s|L^i=l}$ is nothing other than a quantile of the loss distribution $F_s$ of the system $s$ at the level $\tilde{\alpha}=g^{-1}\left(\alpha,F_i\left(l\right)\right)$ i.e. 
\begin{align}
	CoVaR_{\alpha}^{s|L^i=l}=F_s^{-1}\left(\tilde{\alpha}\right).	\label{eq:25}
\end{align}
 Equation \eqref{eq:25} asserts  that $CoVaR_{\alpha}^{s|L^i=l}$ is just a value at risk  of the whole financial system  at a transformed level $\tilde{\alpha}=g^{-1}\left(\alpha,F_i\left(l\right)\right)$. This fact motivates the following corollary, which connects $CoVaR_{\alpha}^{s|L^i=l}$ to the value at risk at the level $\alpha$ of the financial system ($VaR_{\alpha}^s$).
Recall that under assumption~\ref{as}  the value at risk of the system at the level $\tilde{\alpha}$ of $L^s$ is in this case given by 
\begin{align*}
	VaR_{\tilde{\alpha}}^s=F_s^{-1}\left(\tilde{\alpha}\right).
\end{align*}
That is
\begin{align*}
CoVaR_{\alpha}^{s|L^i=l} =	VaR_{\tilde{\alpha}}^s
\end{align*}
\end{rmk}
\begin{co}
\label{co6}
Provided that the   function $g\left(v, u \right):=\frac{\partial C\left(u,v \right)}{\partial u }$ is invertible with respect  to the "non-conditioning" variable $v$, 
 the equivalent confidence level $\tilde{\alpha}$, which makes the Value at Risk of a  financial system $VaR^s$ equivalent to the $CoVaR_{\alpha}^{s|L^i=l}$   at level $\alpha$ is given by:
\begin{align}
	\tilde{\alpha}= g^{-1}\left(\alpha,  u \right)  \ \ with \ u = F_i\left(l\right).
	\label{eq:26}
\end{align}
 \end{co}
Hence, in general, given a condition quantile at the level $\alpha$, we can find the corresponding unconditional quantile by transforming the conditional level $\alpha$ to a unconditional level $\tilde{\alpha}$ through the transformation function $g^{-1}$. Based on the fact that $CoVaR_{\alpha}^{s|L^i=l}$can be expressed as a quantile.  
We can  simplify the expression of  $CoVaR_{\alpha}^{s|L^i=l}$ in a linear function when $L^s$ is assumed to have  a univariate normal distribution. In fact if a random variable $X$ follows a normal distribution with mean $\mu$ and standard deviation $\sigma$.  Then the transformed random variable  $Z=\frac{X-\mu}{\sigma}$ is standard normal distributed. This motivates the  following proposition.
 \begin{prop} 
\label{pr4}
If the loss of the financial system  $L^s$   is assumed to be normal distributed with mean $\mu_s$ and standard deviation $\sigma_s$. Then %$CoVaR_{\alpha}^{s|L^i=l}$ can be compute as follow 
\begin{align}
	CoVaR_{\alpha}^{s|L^i=l}=\sigma_s\Phi^{-1}\left(\tilde{\alpha}\right)+\mu_s,
\end{align}
with  $\tilde{\alpha}$ defined as in equation  \eqref{eq:26}. Where $\Phi$ denotes the standard normal distribution function. 
\end{prop}
That means $CoVaR_{\alpha}^{s|L^i=l}$ is in this case a linear function with respect to the transformation $\Phi^{-1}\left(\tilde{\alpha}\right)$.
\begin{pf}
 Assume that $L^s$ is normal distributed with mean $\mu_s$ and standard deviation $\sigma_s$. Let $N_s$ be the distribution function of  $L^s$ then from \eqref{eq:25} we have 
\begin{align*}
   &CoVaR_{\alpha}^{s|L^i=l}=N_s^{-1}\left(\tilde{\alpha}\right).&\\
 \end{align*}
And using the fact that any arbitrary normal distribution can be transformed  to a standard normal distribution we obtain
\begin{align*}
 & CoVaR_{\alpha}^{s|L^i=l}=\sigma_s\Phi^{-1}\left(\tilde{\alpha}\right)+\mu_s & \  \  \  \  \  \  \blacksquare
\end{align*}
\end{pf}
\begin{co} Under the same conditions as the previous proposition, $ \Delta CoVaR^{s|i}_{\alpha}$ can be compute as follow
\label{c11}
\begin{align}
\label{eq:28}
	 \Delta CoVaR^{s|i}_{\alpha}&=\sigma_s\left(\Phi^{-1}\left(\tilde{\alpha}_d\right)-\Phi^{-1}\left(\tilde{\alpha}_m\right)\right)& 	                       
\end{align}
Where $\tilde{\alpha}_d$ and $\tilde{\alpha}_m$ are   the transformed level defining according to the corollary~\ref{co6} when institution $i$ is under distress  and institution having mean loss respectively.
\end{co}
In the following remark we summarise some properties of  $CoVaR_{\alpha}^{s|L^i=l}$ as monetary measures of risk   with particular  attention to the  concept of coherent risk measures [\cite{coherent} \cite{foellmer} def.~4.5] . This summaries  according to \cite{coherent}, properties that a good risk measure should have.
\begin{rmk}
\label{r6}
As  $CoVaR_{\alpha}^{s|L^i=l}$ can be expressed as quantile of the distribution of the system's loss $F_s$ with respect to the transformed level $\tilde{\alpha}$. It follows that $CoVaR_{\alpha}^{s|L^i=l}$ as a function of $\tilde{\alpha}$ has the same properties like  value at risk as a function of a level $\alpha$.
\end{rmk}
In particular following properties.
\begin{pr}
\begin{itemize}	
\item  $CoVaR_{\alpha}^{s|L^i=l}$   is a coherent measure of risk under elliptical distributions (cf.  \cite{qrm} ex.~6.7).
  \item  $CoVaR_{\alpha}^{s|L^i=l}$ increases when the marginal distribution of the system ($F_s$) has leptokurtosis (heavy-tailed) and positive skewness. (cf. \cite{carol2} § IV.2.8.1). 
  \end{itemize}  
\end{pr}
One important advantage of our formula is that, the expression of  $CoVaR_{\alpha}^{s|L^i=l}$ (see eq.~\eqref{eq:18}) can be separated into two distinct components.
\begin{enumerate}
	\item  On the one hand the marginal distributions $F_i$ and $F_s$, which represent the purely univariate features of  the single financial institution $i$ and the financial system $s$ respectively.
	\item On the other hand the function $g^{-1}$, which represents the dependency structure between the single financial institution $i$ and the system $s$).  
\end{enumerate}
This separation is very important for the analysis of systemic risk property of our formula. First, because it describes how  the systemic contribution of one given financial institution  depends on its marginal distributions $F_i$ and the marginal of the financial system  $F_s$. Secondly, because it allows us to appreciate the effect of the copula of the systemic risk contribution.

%%%%%%%%%%%%%%%%%%%%%%%%%%%%%%%%%%%%%%%%%%%%%%%%%%%%%%%%%%%%%%%%%

%%%%%%%%%%%%%%%%%%%%%%%%%%%%%%%%%%%%%%%%%%%%%%%%%%%%%%%%%%%%%%%%%
 
\section{Tail Events and Systemic Crisis}
\label{sec:4}

As asserted by  \citet{covar}, the main idea of Systemic risk measurement is to  capture the potential for the spreading of financial distress across institutions by gauging the increase in tail comovement (e.g. The prefix $\textsl{Co}$ in \cite{covar}  refers to  conditional, contagion, or comovement).

\begin{mydef}
\citet{forbes} define  contagion as a significant increase in cross-market linkages after a shock to one market (or a group of markets).
\end{mydef}
\begin{rmk}
During the crisis, the contagion effect appears to amplify the concentration of the financial  system  leading to an increase in probability that single financial institutions fail together with the whole financial system or that a large number of financial institutions fail simultaneously.( see Figure~1). 
\end{rmk}
 \begin{figure}[h]
	\begin{center}
		\includegraphics[height=6cm,width=12cm]{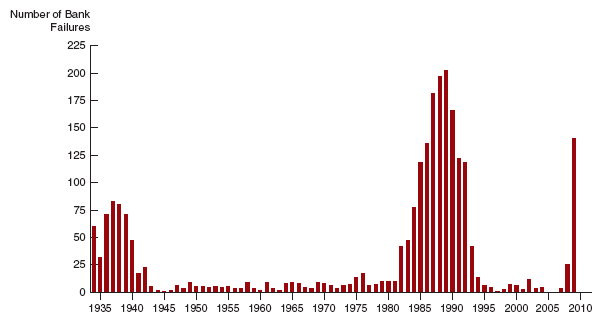}
	\end{center}
	\caption{Bank Failures in the United States, from 1934 to 2009  (Source: \cite{fmi} p.~443)}
\end{figure}
The argumentation above  highlights three important features of systemic risk. 
\begin{enumerate}
	\item Systemic risk involves comovement. 
	\item Systemic risk concerns a precise region of the involved losses distribution (e.g. the tail   by \citet{covar} or the Distress region by \cite{bank_sta} and \cite{qfrm_s}). \citet{carol_b} and \citet{qfrm_s} have taken this fact into consideration and proposed alternative definitions of $CoVaR$ and systemic risk measure respectively. 
	\item Systemic risk involves contagion.
\end{enumerate}
 
Hence in the context of the analysis and the measurement  of systemic risk. The dependence between the  financial institution $i$ and the financial market $s$  have to be considered only in a determined region of their joint distribution. (e.g. in the Tail or in Distress region).

%It is thus clear that an effective Systemic risk measures have to  considers only  the dependence between the focused financial institute $i$ and the financial market $s$  in a determined regions  of their joint distributions (e.g. in the tail or in Distress region). 
One way to do this would be to use  dependence measures which allow the measurement of the dependence only in a defined region(cf. \cite{ext_risk} chap.~6).  
For example in the case where the dependence structure is controlled by the correlation coefficient $\rho$, which is the case for elliptical copulas (e.g. Gaussian and t Copula). The  conditional correlation coefficient has to be used instead of the unconditional correlation coefficient.% , where $A$ refers to the set induced by the condition $C\left(L^i\right)$.
\begin{mydef}[cf. \cite{ext_risk} def. 6.2.1]
Let $U$ and $V$  be two real random variables and  $A$  a subset of $\mathbb{R}$ such that $Pr\left(V \in A\right) > 0$. The conditional correlation coefficient $\rho_A$ of $U$ and $V$ conditioned on $V \in A$ is given by
\begin{align*}
	\rho_A=\frac{Cov\left(U,V| V\in A\right)}{\sqrt{Var\left(U|V\in A\right)\cdot Var\left(V|V\in A\right)}}.
\end{align*}
\end{mydef} 

Recall that the condition $C\left(L^i\right)$ in \cite{covar} refers to the loss  being exactly at some specific
value (e.g. $VaR_i$) and because of assumption~1 we have that $Pr\left(L^i=l\right)=0$ for any $l\in \mathbb{R}$. To circumvent this problem we proceed as follows. Instead of considering the set where $L^i$  is assume some fixed value  we follows \citet{feller} p.~71  and consider the set where $L^i$ assumes values in an interval $I=\left(l, l+\Delta l\right)$.
We define 
\begin{align}
	\rho_{=}:=\lim_{\Delta l \to 0}\frac{Cov\left(L^s,L^i| L^i\in I\right)}{\sqrt{Var\left(L^s|L^i\in I\right)\cdot Var\left(L^i|L^i\in I\right)}}.
\end{align}
  \begin{rmk}
 The use of the conditional correlation coefficient instead the unconditional allows the investigation  of the effect of the Condition $C\left(L^i\right)$ on the systemic risk contribution.
 \end{rmk}
 
 Another tool to measure  the dependence of two random variables  in one precise given region of their joint  distribution  is  the so called quantile-quantile dependence measure $\lambda\left(\alpha\right)$ introduced by cf. \cite{coles}. This is defined as   % as defined by \citet{ronc2} (cf. \cite{ronc2} Remarque~58). 
\begin{align*}
	 	\lambda_u\left(\alpha\right)=Pr \left(V > G^{-1}\left(\alpha\right) | U > F^{-1}\left(\alpha\right) \right).
\end{align*}
So according to the previous argumentation. The quantile-quantile dependence measure $\lambda\left(\alpha\right)$  of $L^i$ and $L^s$ is  per definition a natural indicator of the contagion between financial institutions over a threshold $\alpha$ (Note that, the typical value of $\alpha$ in our context are  0.99 or 0.995 ).  $\lambda_u(\alpha)= 0$, for example, could  mean that there is no contagion between $i$ and $s$ over the threshold $\alpha$ ).

\begin{rmk} 
Let us consider a "lower-version" of the quantile-quantile dependence measure as defined by \citet{ronc2} (cf. \cite{ronc2} Remarque~58). 
\begin{align*}
	 	\lambda_l\left(\alpha\right)=Pr \left(V < G^{-1}\left(\alpha\right) | U < F^{-1}\left(\alpha\right) \right).
\end{align*}
If we redefine the condition $C\left(L^i\right)$ in the implicit definition of $CoVaR_{\alpha}^{s|C\left(L^i\right)}$ (see. eq.\eqref{eq:3} and  \eqref{eq:4}) by replacing "=" by "$\leq$" we have the interesting relation.
\begin{align*}
	Pr\left(L^s\leq CoVaR_{\alpha}^{s|C\left(L^i\right)}|C\left(L^i\right)\right)&=Pr\left(L^s\leq CoVaR_{\alpha}^{s|C\left(L^i\right)}|L^i\leq l\right)&\\
	 ''\text{Assume that } CoVaR_{\alpha}^{s|C\left(L^i\right)}=F_s^{-1}\left(\hat{\alpha}\right)''    &=Pr\left(L^s\leq F_s^{-1}\left(\hat{\alpha}\right)|L^i\leq F_i^{-1}\left(\beta\right)\right)& \\
	       ''\text{Assume that  $\hat{\alpha}=\beta$ then we have }''  &=Pr\left(L^s\leq F_s^{-1}\left(\beta\right)|L^i\leq F_i^{-1}\left(\beta\right)\right)& \\  
	                                                                            &=\lambda_l\left(\beta\right).&                                                                    
\end{align*}
\end{rmk}
 The asymptotic consideration of $\lambda_u\left(\alpha\right)$ and $\lambda_l\left(\alpha\right)$  leads to the following definitions.
 
\begin{mydef}[cf. \cite{qrm} def.~5.30] Let $\left(U,V\right)$ be a bivariate random variable with marginal distribution functions $F$ and $G$, respectively. The upper tail dependence coefficient of $U$ and $V$ is   the limit (if it exists) of the conditional probability that
$V$ is greater than the $100 \alpha-th$ percentile of $G$ given that $U$ is greater than
the $100\alpha-th$ percentile of $F$ as $\alpha$ approaches $1$, i.e.
\begin{align*}
	\lambda_u:=\lim_{\alpha \to 1^-}{\lambda_u\left(\alpha\right)}=\lim_{\alpha \to 1^-}Pr \left(V > G^{-1}\left(\alpha \right) | U > F^{-1}\left(\alpha\right) \right)	                   
\end{align*}
If $\lambda_u \in $  (0, 1] then $\left(U,V\right)$ is said to show
upper tail dependence or extremal dependence in the upper tail; if $\lambda_u =0$, they are
asymptotically independent in the upper tail.\\
Similarly, the lower tail dependence coefficient $\lambda_l$ is the limit (if it exists)
of the conditional probability that $V$ is less than or equal to the
$100\alpha-th$ percentile of $G$ given that $U$ is less than or equal to the $100\alpha-th$
percentile of $F$ as $\alpha$ approaches 0, i.e.
\begin{align*}
	\lambda_l:=\lim_{\alpha \to 0^+}{\lambda_l\left(\alpha\right)}=\lim_{\alpha \to 0^+}Pr \left(V \leq G^{-1}\left(\alpha\right) | U \leq F^{-1}\left(\alpha\right) \right).
\end{align*}
\end{mydef}

If $\left(L^i,L^s\right)$ does not show tail dependence  (upper and lower) the extreme events of $L^i$ and $L^s$  appear to occur independently  in each margin. This means that they are no-contagion betwenn $i$ and $s$.

Let us consider the  bivariate Gaussian copula as model for $\left(L^i,L^s\right)$. One can show that the bivariate Gaussian copula does not have upper tail dependence when the corresponding  correlation coefficient $\rho$ is smaller than one  (see  \ref{tail_g}). As can be seen in Figure~2, regardless of how high a correlation we choose, if we go far enough into the tail, extreme events appear to occur independently in $L^i$ and $L^s$. That means the Gaussian copula is related to the independence in the tail. Thus  Gaussian copula is not a good model for the analysis of  systemic risk contribution between $i$ and $s$. This is the reason why we connect the $CoVaR$ concept to copula in order to develop a closed formula for $CoVaR^{s|L^i=l}$  allowing the analysis and the computation of systemic risk contribution for a more general stochastic setting  than only the bivariate Gaussian setting as already done in (\cite{manfred}). %So we have to explore alternative  
 
%%%%%%%%%%%%%%%%%%%%%%%%%%%%%%%%%%%%%%%%%%%%%%%%%%%%%%%%%%%%%%%%%

%%%%%%%%%%%%%%%%%%%%%%%%%%%%%%%%%%%%%%%%%%%%%%%%%%%%%%%%%%%%%%%%%
 
\section{Applications}
In this section we apply the result developed  in section \ref{sec:3} to compute and analyse  $CoVaR_{\alpha}^{s|L^i=l}$ and $\Delta CoVaR$ in some probabilistic settings.  
We first consider a general case where the joint behavior of  $L^i$ and $L^s$ is modeled by a bivariate Gaussian copula. In particular we will analyse here the case where the margins $L^i$  and  $L^s$ are assumed to be univariate normal distributed. This special  case (Gauss copula and Gaussian margnis) was already considered in \cite{manfred} but in a different approach.   \citet{manfred} assumes that the random vector $\left(L^i L^s\right)$ follows a bivariate  Gaussian distribution. Then, based on the properties of the conditional  bivariate Gaussian distribution  (cf. e.g. \cite{feller} eq.~2.6), he develops  a closed formula for $CoVaR_{\alpha}^{s|L^i=VaR_i}$. This approach imposes thus  the univariate normality of both margins ($L^i$   and $L^s $). The method provided in this article is more flexible because it allows   each margins independently of other  to take a large class of distributions functions (for example we ca assume that $L^i$ is normal distributed and that $L^s$ is $t$-distributed). One other restriction of the formula proposed in \cite{manfred} and also the method presented in \cite{covar} is that, both do not take into account tail events and tail comovements since the  Gaussian is asymptotically  independent in both tails i.e. $\lambda_u=\lambda_l =0$ . In fact we have  (cf.   \cite{emb} p.~17). 
\begin{align}
\label{tail_g}
	\lambda_u&=2 \lim_{\alpha \to \infty}\left[1- \Phi\left(\frac{\alpha-\rho \alpha}{\sqrt{1-\rho^2}}\right) \right]& \nonumber\\
	       &=2 \lim_{\alpha \to \infty}\left[1- \Phi\left(\frac{\alpha\sqrt{1-\rho}}{\sqrt{1+\rho}}\right) \right],&
\end{align}
from which it follows that 
\begin{align*}
\lambda_u	 = \left\{ \begin{array}{rl}
  0 &\mbox{ if $\rho < 1$} \\
  1 &\mbox{ if $\rho = 1$.}
       \end{array} \right.
\end{align*}
  \begin{figure}[h]
  \label{f_tail_g}
	\begin{center}
		\includegraphics[height=6cm,width=10cm]{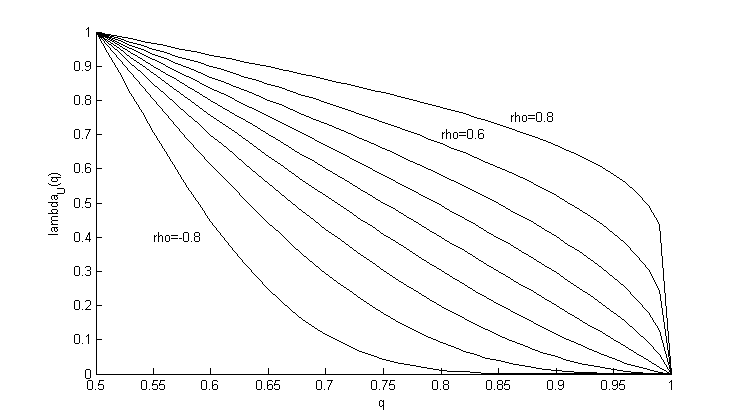}
	\end{center}
	\caption{ $\lambda _u\left(\alpha\right)$ for bivariate Gaussian Copula}
\end{figure} This presents a big gap since both phenomenons (tail events and tail comovements) are supposed to be the main features   of systemic crisis (cf. \cite{acharya1}). Our formula covers this gap by  allowing  us to consider other  dependence models, especially those which are appropriate for the modeling of the simultaneous (tail) behavior of losses during a financial crisis.  So we will also consider the case where the dependence between $L^i$ and $L^s$ is modeled  by a $t$-Copula, Gumbel-Copula. At the end of this section we will describe how to develop a closed formula for the computation of $CoVaR_{\alpha}^{s|L^i=l}$ for  Archimedean copula.
\subsection{Computation of $CoVaR_{\alpha}^{s|L^i=l}$ in a Gaussian Copula Setting}
\label{ex1}
We assume here that the interdependence structure between $L^i$ and $L^s$ is describe by a bivariate Gaussian copula. The bivariate Gaussian copula is defined as follows (cf. \cite{nelsen} eq.~2.3.6 ):
\begin{align*}
	C_{\rho}\left(u,v\right)=\Phi_2\left(\Phi\left(u\right)^{-1},\Phi\left(v\right)^{-1}\right),
\end{align*}
where  $\Phi_2$ denotes the bivariate standard normal distribution with linear correlation coefficient  $\rho$, and $\Phi$ the univariate standard normal  distribution.\\
Hence,
\begin{align*}
	C_{\rho}\left(u,v\right)=\int_{-\infty}^{\Phi^{-1}\left(u\right)}\int_{-\infty}^{\Phi^{-1}\left(v\right)}\frac{1}{2\pi \sqrt{1-\rho^2}}\exp\left(\frac{2\rho st-s^2-t^2}{2\left(1-\rho^2\right)}\right)dsdt.
\end{align*}
 Note that   $C_{\rho}\left(u,v\right)$ can be express as  
\begin{align}
\label{eq_29}
	C_{\rho}\left(u,v\right)=\int_0^u \Phi\left(\frac{\Phi^{-1}\left(v\right)-\rho\Phi^{-1}\left(t\right)}{\sqrt{1-\rho^2}}\right)dt.
\end{align}
 
In fact let $X = (U, V)$ a standard Gaussian random vector with correlation $\rho$. Then we have:
\begin{align*}
	\Phi_2\left(u,v \right)&=Pr \left(U\leq u, V\leq v \right)&\\
	&=\int_{-\infty}^{u}\int_{-\infty}^{v}\frac{1}{2\pi \sqrt{1-\rho^2}}exp\left(\frac{2\rho st-s^2-t^2}{2\left(1-\rho^2\right)}\right)dsdt&
\end{align*}
this implies that
\begin{align*}
	\frac{\partial \Phi_2\left(u,v\right)}{\partial u }& 	= \int_{-\infty}^{v}\frac{1}{2\pi \sqrt{1-\rho^2}}exp\left(\frac{2\rho ut-u^2-s^2}{2\left(1-\rho^2\right)}\right)ds& \nonumber\\
	       & 	= \int_{-\infty}^{v}\frac{1}{2\pi \sqrt{1-\rho^2}}exp\left(\frac{-\left(s-u\rho\right)^2 +\rho^2u^2 - u^2}{2\left(1-\rho^2\right)}\right)ds& \nonumber\\
	       & 	= \int_{-\infty}^{v}\frac{1}{2\pi \sqrt{1-\rho^2}}exp\left( \frac{-\left(s-u\rho\right)^2 - u^2\left(1-\rho^2\right)  }{2\left(1-\rho^2\right)}\right)ds& \nonumber\\
	       & 	= \int_{-\infty}^{v}\frac{1}{2\pi \sqrt{1-\rho^2}}exp\left( \frac{ - u^2 }  {2} + \frac{-\left(s-u\rho\right)^2   }  {2\left(1-\rho^2\right)}\right)ds& \nonumber\\
	       & 	= \frac{1}{ \sqrt{2\pi}}exp\left( \frac{ - u^2 }{2} \right)\int_{-\infty}^{v}\frac{1}{\sqrt{2\pi} \sqrt{1-\rho^2}}exp\left( \frac{-\left(s-u\rho\right)^2   } 
 {2\left(1-\rho^2\right)}\right)ds& \nonumber\\
  &=\phi\left(u\right)\cdot\Phi\left(\frac{v-u\rho }{\sqrt{1-\rho^2}}\right),&
\end{align*}
where $\phi$ denotes the density of the standard univariate normal distribution.
Therefore we have,  
\begin{align*}
	\Phi_2\left(u,v\right)& 	= \int_{-\infty}^{u}\phi\left(x\right)\cdot\Phi\left(\frac{v-x\rho }{\sqrt{1-\rho^2}}\right)dx,&
\end{align*}
The expression of the bivariate Gaussian copula is then
\begin{align*}
	C_{\rho}\left(u,v\right)&=	\Phi_2\left(\Phi^{-1}\left(u\right),\Phi^{-1}\left(v\right),\rho \right)&\\
	                     &=\int_{-\infty}^{\Phi^{-1}\left(u\right)}\phi\left(x\right)\cdot\Phi\left(\frac{\Phi^{-1}\left(v\right)-x\rho }{\sqrt{1-\rho^2}}\right)dx&
\end{align*}
By making  the substitution $t = \Phi\left(x\right)$, we obtain
\begin{align*}
	&C_{\rho}\left(u,v\right)=\int_0^u \Phi\left(\frac{\Phi^{-1}\left(v\right)-\rho\Phi^{-1}\left(t\right)}{\sqrt{1-\rho^2}}\right)dt. 
\end{align*} 
 
By considering the expression \eqref{eq_29} we have:
\begin{align}
  g\left(v, u\right) &=\frac{\partial C_{\rho}\left(u,v \right)}{\partial u }& \nonumber\\
	&=\frac{\partial\left( \int_0^u \Phi\left(\frac{\Phi^{-1}\left(v\right)-\rho\Phi^{-1}\left(t\right)}{\sqrt{1-\rho^2}}\right)dt \right)}{\partial u }& \nonumber\\
&	=\Phi\left(\frac{\Phi^{-1}\left(v\right)-\rho\Phi^{-1}\left(u\right)}{\sqrt{1-\rho^2}}\right)&
 \end{align}
 %\begin{figure}[h]
	%\begin{center}
%	\label{f2}
		%\includegraphics[height=4.5cm,width=12cm]{g}
%	\end{center}
	%\caption{ Conditional Quantile Function of the Bivariate Gaussian Copula with $\rho = 0.5$}
%\end{figure} 
The function $g\left(v, u\right)$ is strictly monotone with respect to $v$. 
\begin{figure}[h]
\label{f3}
	\begin{center}
		\includegraphics[height=4.5cm,width=12cm]{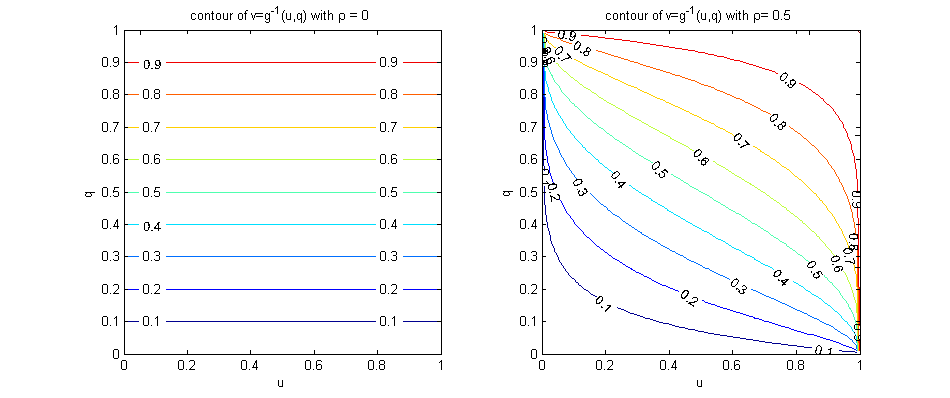}
	\end{center}
	\caption{$g^{-1}$ of the Bivariate Gaussian Copula }%$v=g^{-1}_{q,\rho}\left(u\right)$}
\end{figure}
To compute its inverse, we set $g\left(v,u\right)=\alpha $ and solve for $v$.  
 \begin{align*}
v=g^{-1}\left(\alpha, u\right)=\Phi\left(\rho\Phi^{-1}\left(u\right)+\sqrt{1-\rho^2}\Phi^{-1}\left(\alpha\right)\right).
\end{align*}
\begin{rmk}If we set in the above equation $\rho = 0$, we obtain $v=\alpha$ for all  $u \in \left[0, 1\right]$ (see Figure~\ref{f3})). This is not a surprise because  zero correlation means independence under the normal copula setting.
\end{rmk}
So according to theorem \ref{main} and the development make in section 4 we have the following formula for $CoVaR_{\alpha}^{s|L^i=l}$ when the dependence is modeled by a Gaussian copula.

\begin{prop}
Assume that the copula of $L^i$ and  $L^s$ is the Gaussian copula then
\begin{align}
\label{eq:36}
	CoVaR^{s|L^i=l}=F^{-1}_{s}\left(\Phi\left(\rho_{=}\Phi^{-1}\left(F_{i}\left(l\right)\right)+\sqrt{1-\rho_{=}^2}\Phi^{-1}\left(\alpha\right)\right)\right).
\end{align}
In the context of remark \ref{r2}, we have
\begin{align*}
	\tilde{\alpha}=\Phi\left(\rho_{=}\Phi^{-1}\left(F_{i}\left(l\right)\right)+\sqrt{1-\rho_{=}^2}\Phi^{-1}\left(\alpha\right)\right),
\end{align*}

Where $F^i$ and $F^s$ represent the univariate distribution function of $L^i$ and $L^s$ respectively. 
\end{prop}

%Note that we can be $F_i$ and $F_s$ in \eqref{eq:36} any univariate distribution functions satisfying  assumption~\ref{as}. \\

Let us suppose as a particular case that $L^i$ and $L^s$ are  Gaussian, that is $F_s =N_{s}$  and  $F_i =N_{i}\ $,  where $N_{i}$ and $N_{s}$ are Gaussian distributions of the losses $L^i$ and $L^s$ with expected values $\mu_i$, $\mu_s$ and standard deviation $\sigma_s$, $\sigma_i$ (This correspond to the case considered  in \cite{manfred}).  This h
We obtain the following  closed analytical expression of $CoVaR^{s|L^i=l}$% that is :% $u=F_{L^i}\left(VaR^i_q\right)$ and $v=F_{L^s}\left(CoVaR^{s|L^i=E(L^i)}\right)$
%\begin{align}
%		CoVaR^{s|L^i=l}=N^{-1}_{s}\left(\Phi\left(\rho_{=}\Phi^{-1}\left(N_{i}\left(l\right)\right)+\sqrt{1-\rho_{=}^2}\Phi^{-1}\left(\alpha\right)\right)\right)
%\end{align}
%and 
%\begin{align}
%\label{eq:47}
%		\tilde{\alpha}=\Phi\left(\rho_{=}\Phi^{-1}\left(N_{i}\left(l\right)\right)+\sqrt{1-\rho_{=}^2}\Phi^{-1}\left(\alpha\right)\right).
%\end{align}
in the Gaussian setting (Gaussian Copula and Gaussian Margins) 
 \begin{align}
\label{eq:48}
	CoVaR^{s|L^i=l} = \rho_{=}\frac{\sigma_s}{\sigma_i} \left( l   -  \mu_i \right) + \sqrt{1-\rho_{=}^2} \sigma_s\Phi^{-1}\left(\alpha\right)+\mu_s. 
\end{align}
 
In fact we have
 \begin{align}
		CoVaR^{s|L^i=l}&=N^{-1}_{s}\left(\Phi\left(\rho_{=}\Phi^{-1}\left(N_{i}\left(l\right)\right)+\sqrt{1-\rho_{=}^2}\Phi^{-1}\left(\alpha\right)\right)\right)& \nonumber \\
		          &=N^{-1}_{s}\left(N_{s}\left(\sigma_s\rho_{=}\Phi^{-1}\left(N_{i}\left(l\right)\right)+\sigma_s\sqrt{1-\rho_{=}^2}\Phi^{-1}\left(\alpha\right)+\mu_s\right)\right)& \nonumber\\ 
		           &=\sigma_s\rho_{=}\Phi^{-1}\left(N_{i}\left(l\right)\right)+\sigma_s \sqrt{1-\rho_{=}^2}\Phi^{-1}\left(\alpha\right)+\mu_s& \nonumber \\
		           &=\sigma_s\rho_{=}\Phi^{-1}\left(\Phi\left(\frac{l-\mu_i}{\sigma_i}\right)\right)+\sigma_s\sqrt{1-\rho_{=}^2}\Phi^{-1}\left(\alpha\right)+\mu_s & \nonumber \\
		            &=\sigma_s\rho_{=}\left(\frac{l-\mu_{i}}{\sigma_i}\right)+\sigma_s\sqrt{1-\rho_{=}^2}\Phi^{-1}\left(\alpha\right)+\mu_s& \nonumber\\
		       %     &=\sigma_s\rho_{=}\frac{l}{\sigma_i}   -  \sigma_s\rho_{=}\frac{\mu_i}{\sigma_i}  +\sigma_s\sqrt{1-\rho_{=}^2}\Phi^{-1}\left(\alpha\right)+\mu_s& \nonumber \\  
		       %     &=\rho_{=}\frac{\sigma_s}{\sigma_i}l   -  \rho_{=}\frac{\sigma_s}{\sigma_i}\mu_i +\sigma_s\sqrt{1-\rho_{=}^2}\Phi^{-1}\left(\alpha\right)+\mu_s & \nonumber\\
		        %    &=\rho_{=}\frac{\sigma_s}{\sigma_i}l   -  \rho_{=}\frac{\sigma_s}{\sigma_i}\mu_i  +\sqrt{1-\rho_{=}^2} \sigma_s\Phi^{-1}\left(\alpha\right)+\mu_s & \nonumber\\
		            &=\rho_{=}\frac{\sigma_s}{\sigma_i} \left( l   -  \mu_i \right) + \sqrt{1-\rho_{=}^2} \sigma_s\Phi^{-1}\left(\alpha\right)+\mu_s. \nonumber		            
\end{align}
 And we have in this case
 (cf. \cite{ext_risk} eq.~6.1)
\begin{align}
	\rho_= =\lim_{\Delta l \to 0^+}\frac{\rho}{\sqrt{\rho^2+\left(1-\rho^2\right)\frac{Var\left(L^i\right)}{Var\left(L^i|L^i\in I\right)}}}
\end{align}
\begin{rmk}
Note that $\rho_=$ can be either greater or smaller than $\rho$ since $Var\left(L^i|L^i\in I\right)$ can be either greater
or smaller than $Var\left(L^i\right)$. This fact have to be considered when analysing the effect of the dependence parameter $\rho$ on   $CoVaR$. The consideration of $\rho_=$  highlights also the impact that the change in volatility can have on the systemic risk contribution. This a very important since the behavior of the volatility are not the same depending if we are in distress region or not.  In fact one can observe that  distress times are  characterised by high volatility.
\end{rmk}
 
 \begin{co} 
Assume that $L^s$ and $L^i$ are Gaussian distributed and  centered at zero then.
 \begin{align*}
	CoVaR^{s|L^i=l}% &= \rho_{=}\frac{\sigma_s}{\sigma_i} \left( l   -  \mu_i \right) + \sqrt{1-\rho_{=}^2} \sigma_s\Phi^{-1}\left(\alpha\right)+\mu_s,& \nonumber \\
	                  &= \left(\rho_{=}\frac{\sigma_s}{\sigma_i}\right)   l    +  \sqrt{1-\rho_{=}^2} \sigma_s\Phi^{-1}\left(\alpha\right)& \nonumber\\
	                   &= \left(\rho_{=}\frac{\sigma_s}{\sigma_i}\right)VaR_{\beta}^i +  \sqrt{1-\rho_{=}^2} VaR^s_{\alpha} \  \   \   \   when \   l=VaR_{\beta}^i  
\end{align*}
\end{co}
\begin{rmk}
 If $\rho_{=} = 0$  then  $CoVaR^{s|L^i=VaR^i}=VaR^s_{\alpha}.$
\end{rmk} 
 
Let $l$ be the value at risk of the single institution at the level $\beta$ i.e. $l= VaR^i_\beta=F_i\left(\beta\right)$. Then we have the following expression of $CoVaR_{\alpha}^{\beta}$ (see def.~ \ref{d6}). 

\begin{co}In the Gaussian setting, we have %the $CoVaR^{s|L^i=VaR^i_\beta}$ is :  
 \begin{align}
 \label{eq:49}
 CoVaR_{\alpha}^{\beta}  =\rho_{=}\sigma_s\Phi^{-1}(\beta)  + \sqrt{1-\rho_{=}^2} \sigma_s\Phi^{-1}\left(\alpha\right)+\mu_s.
   \end{align}
 \end{co} 
  \begin{pf}   
   \begin{align*}
	CoVaR_{\alpha}^{\beta}  &= h\cdot VaR^i_\beta -h \cdot \mu_i +\sqrt{1-\rho_{=}^2} \sigma_s\Phi^{-1}\left(\alpha\right)+\mu_s.& \\
	                    &=\rho_{=}\frac{\sigma_s}{\sigma_i}\left(\sigma_i\Phi^{-1}(\beta)+\mu_i\right) - \rho_{=}\frac{\sigma_s}{\sigma_i} \mu_i + \sqrt{1-\rho_{=}^2} \sigma_s\Phi^{-1}\left(\alpha\right)+\mu_s \nonumber &\\
	                    &=\rho_{=}\sigma_s\Phi^{-1}(\beta)  +\sqrt{1-\rho_{=}^2} \sigma_s\Phi^{-1}\left(\alpha\right)+\mu_s.  \nonumber  &
	                      & &\blacksquare  \nonumber
   \end{align*}
  \end{pf}
  \begin{rmk} We remark that unlike in equation ~\ref{eq:48} the expression of $CoVaR_{\alpha}^{\beta} $ does not depend of the loss distribution's characteristic (e.g. standard deviation $\sigma_i$ and mean $\mu_i$) of the financial institution $i$. 
  \end{rmk}
  \begin{co}
In the Gaussian setting. The map 
\begin{align*} 
	\left(\alpha, \beta\right) \longmapsto CoVaR_{\alpha}^{\beta} 
\end{align*}
 is increasing with respect to its both parameters.
  \end{co}
  Now we refer to definition \ref{d4} to compute $\Delta CoVaR^{s|i}_{\alpha}$. The result of our computation is provided in the  
following proposition.% provides a formula for the computation of  $\Delta CoVaR^{s|i}_{\alpha}$ in a Gaussian setting. 
\begin{prop}
 In the Gaussian setting, $\Delta CoVaR^{s|i}_{\alpha}$ is given by
\begin{align}
\label{eq:49}
\Delta CoVaR^{s|i}_{\alpha} =   \rho_{=}\sigma_s\Phi\left(\beta\right)^{-1}.  
\end{align}
\end{prop}
\begin{pf}According to  definition \ref{d4} we have
\begin{align*}
\Delta CoVaR_{\alpha}^{\beta} &= CoVaR^{s|L^i=VaR^i_{\beta}}_{\alpha} - CoVaR^{s|L^i = \mu_{i} }_{\alpha} \nonumber &\\
                           &=   \rho_{=}\frac{\sigma_s}{\sigma_i}\cdot VaR^i_{\beta} -\rho_{=}\frac{\sigma_s}{\sigma_i} \cdot \mu_{i} + \sqrt{1-\rho_{=}^2} VaR^s_{\alpha} - \left[ \rho_{=}\frac{\sigma_s}{\sigma_i}\cdot  \mu_{i}  - \rho_{=}\frac{\sigma_s}{\sigma_i} \cdot \mu_{i} + \sqrt{1-\rho_{=}^2} VaR^s_{\alpha} \right]&  \nonumber  \\ 
                           &=   \rho_{=}\frac{\sigma_s}{\sigma_i}\cdot VaR^i_{\beta}    -   \rho_{=}\frac{\sigma_s}{\sigma_i}\cdot  \mu_{i}      & \nonumber  \\
                              &=   \rho_{=}\frac{\sigma_s}{\sigma_i}\cdot \left(VaR^i_{\beta}    -   \mu_{i}\right)   &  \nonumber \\
                           &=   \rho_{=}\frac{\sigma_s}{\sigma_i}\cdot \left( \sigma_i \Phi^{-1}\left( \beta \right)+ \mu_{i}    -     \mu_i\right)  & \nonumber \\
                             &=   \rho_{=}\sigma_s\Phi\left(\beta\right)^{-1}. &     \blacksquare                                                            
\end{align*}
\end{pf}
\begin{rmk} From equation \eqref{eq:49} we observe that if the financial institution $i$ and the financial system $s $ are not correlated, the risk contribution of  $i$ to $s$ is zero. 
\end{rmk}
Let us impose now that, the loss of the financial system $L^s$ alone follows  normal univariate distribution. Then according to proposition~\ref{pr4}, we have
\begin{align*}
	CoVaR_{\alpha}^{s|L^i=l}=\sigma_s \Phi^{-1}\left(\tilde{\alpha}\right)+\mu_s.                    
\end{align*}
Additionally if we also assume that $L^i$ is normal distributed. Then we have   
\begin{align*}
	\tilde{\alpha}&=\Phi\left(\rho_{=}\Phi^{-1}\left(N_i\left(l\right)\right)+\sqrt{1-\rho_{=}^2}\Phi^{-1}\left(\alpha\right)\right)&\\
	              &=\Phi\left(\rho_{=}\Phi^{-1}\left(\beta\right)+\sqrt{1-\rho_{=}^2}\Phi^{-1}\left(\alpha\right)\right).&
\end{align*}
In sum we have
\begin{align*}
	CoVaR_{\alpha}^{s|L^i=l} &	=\sigma_s \left(\Phi^{-1}\left(\Phi \left(\rho_{=}\Phi^{-1}\left(N_i\left(l\right)\right)+\sqrt{1-\rho_{=}^2}\Phi^{-1}\left(\alpha\right)\right)\right)\right)+\mu_s& \nonumber \\
		&	=\sigma_s \left(\rho_{=}\Phi^{-1}\left(N_i\left(l\right)\right)+\sqrt{1-\rho_{=}^2}\Phi^{-1}\left(\alpha\right)\right)+\mu_s& \nonumber \\
		&	=\sigma_s \left(\rho_{=}\Phi^{-1}\left( \Phi\left(\frac{l-\mu_i}{\sigma_i}\right)\right)+\sqrt{1-\rho_{=}^2}\Phi^{-1}\left(\alpha\right)\right)+\mu_s&  \nonumber\\
	 	&	=\sigma_s \left(\rho_{=}\left(\frac{l-\mu_i}{\sigma_i}\right)+\sqrt{1-\rho_{=}^2}\Phi^{-1}\left(\alpha\right)\right)+\mu_s&  \nonumber\\
	 	&	=\rho_{=}\frac{\sigma_s }{\sigma_i}\left(l-\mu_i\right)+   \sigma_s\sqrt{1-\rho_{=}^2}\Phi^{-1}\left(\alpha\right)+\mu_s& %\label{eq:51}
\end{align*}		
and 
\begin{align*}
			CoVaR_{\alpha}^{\beta}	&	=\sigma_s \left(\Phi^{-1}\left(\Phi \left(\rho_{=}\Phi^{-1}\left(\beta\right)+\sqrt{1-\rho_{=}^2}\Phi^{-1}\left(\alpha\right)\right)\right)\right)+\mu_s& \nonumber \\
		%&	=\sigma_s \left(\rho_{=}\Phi^{-1}\left(\beta\right)+\sqrt{1-\rho_{=}^2}\Phi^{-1}\left(\alpha\right)\right)+\mu_s& \nonumber \\
	  &	=\sigma_s \left(\rho_{=}\Phi^{-1}\left(\beta\right)+\sqrt{1-\rho_{=}^2}\Phi^{-1}\left(\alpha\right)\right)+\mu_s& \nonumber \\
	  &	=\sigma_s \rho_{=}\Phi^{-1}\left(\beta\right)+\sigma_s \sqrt{1-\rho_{=}^2}\Phi^{-1}\left(\alpha\right) +\mu_s.& %\label{eq:52}
\end{align*}
Similary   we can compute $\Delta CoVaR^{s|i}_{\alpha}$ as follows. Recall (see corollary~\ref{c11})
\begin{align*}
\label{eq:53}
	 \Delta CoVaR^{s|i}_{\alpha}= \sigma_s\left( \Phi^{-1}\left(\tilde{\alpha_d}\right)-\Phi^{-1}\left(\tilde{\alpha_m}\right)\right).                       
\end{align*}
And we have 
\begin{align*}
	\tilde{\alpha}_m& =\Phi\left(\rho_{=}\Phi^{-1}\left(0.5\right)+\sqrt{1-\rho_{=}^2}\Phi^{-1}\left(\alpha\right)\right)&\\
	                  & =\Phi\left(\sqrt{1-\rho_{=}^2}\Phi^{-1}\left(\alpha\right)\right)&
\end{align*}
and 
\begin{align*}
	\tilde{\alpha}_d& =\Phi\left(\rho_{=}\Phi^{-1}\left(\beta\right)+\sqrt{1-\rho_{=}^2}\Phi^{-1}\left(\alpha\right)\right).&
\end{align*}
Hence
\begin{align*}
	 \Delta CoVaR^{s|i}_{\alpha}& = \sigma_s\left( \Phi^{-1}\left(\Phi\left(\rho_{=}\Phi^{-1}\left(\beta\right)+\sqrt{1-\rho_{=}^2}\Phi^{-1}\left(\alpha\right)\right)\right)-\Phi^{-1}\left(\Phi\left(\sqrt{1-\rho_{=}^2}\Phi^{-1}\left(\alpha\right)\right)\right)\right)& \nonumber \\       
	     &=\sigma_s\left( \left(\rho_{=}\Phi^{-1}\left(\beta\right)+\sqrt{1-\rho_{=}^2}\Phi^{-1}\left(\alpha\right)\right) -  \left(\sqrt{1-\rho_{=}^2}\Phi^{-1}\left(\alpha\right)\right)\right) &\nonumber \\                   &=\sigma_s\rho_{=}\Phi^{-1}\left(\beta\right). &
\end{align*}

%%%%%%%%%%%%%%%%%%%%%%%%%%%%%%%%%%%%%%%%%%%%%%%%%%%%%%

%%%%%%%%%%%%%%%%%%%%%%%%%%%%%%%%%%%%%%%%%%%%%%%%%%%%%%%%%%
\subsection{t-copula}  
\label{ex2}
 The Student t copula represents a generalization of the normal copula by allowing for tail-dependence through the degrees of
freedom parameter.

\begin{mydef}[bivariate t distribution]
The  distribution function  of a bivariate  t-distributed random variable with correlation coefficient $\rho$ is given by:
\begin{align*}
	t_{\rho,\nu}\left(u,v\right)=\int_{-\infty}^{u}\int_{-\infty}^{v}\frac{ 1}{2\pi\sqrt{1-\left(\rho\right)^2}}\left(1+\frac{s^2+t^2-2\rho^t st}{ \nu \left(1-\left(\rho\right)^2\right)}\right)^{-\frac{\nu+2}{2}} dsdt,
\end{align*}
where $\nu$ is the number of degrees of freedom.  
\end{mydef}

The Student t copula can be consider as  a generalization of the Gaussian copula. He has in addition to correlation coefficient $\rho$  a second dependence parameter, the degree of freedom $\nu$, controls the heaviness of the tails. For $\nu < 3$, the variance does not exist and for $\nu < 5$, the fourth moment does not exist. The  t copula and the the Gaussian copula are close to each other in their central part, and become closer and closer in their tail only when $\nu$ increases. Especially for both copulas are almost identic when $\nu \rightarrow \infty$ 
freedom parameter.

 \begin{mydef}
The bivariate t copula, $C^t_{\rho,\nu}$ is defined as
\begin{align*}
	C^t_{\rho,\nu}\left(u,v\right)&=t_{\rho,\nu}\left(t_\nu^{-1}\left(u\right),t_\nu^{-1}\left(v\right)\right)&\\
	&=\int_{-\infty}^{t_\nu^{-1}\left(u\right)}\int_{-\infty}^{t_\nu^{-1}\left(v\right)}\frac{ 1}{2\pi\sqrt{1-\left(\rho\right)^2}}\left(1+\frac{s^2+t^2-2\rho^t st}{ \nu \left(1-\left(\rho\right)^2\right)}\right)^{-\frac{\nu+2}{2}} ds dt.&
\end{align*}
\end{mydef}
The tail-dependence coefficients the t Copula is given by (cf. e.g. \cite{qrm} eq.~(5.31))
Because of the symmetric property of $t$ distribution we have,
\begin{align*}
	\lambda_l=\lambda_u =2 -2 t_{\nu+1}\left(\left(\frac{\left(\nu+1\right)\left(1-\rho\right)}{1+\rho}\right)^{\frac{1}{2}}\right).&
\end{align*}
From which it follows that,
\begin{align*}
\lambda_u	 = \left\{ \begin{array}{rl}
 > 0 &\mbox{ if $\rho > - 1$} \\
  0 &\mbox{ if $\rho = -1$ .}
       \end{array} \right.
\end{align*}
Provided that $\rho > 1$. The  bivariate $t$ copula is thus  able to capture the  dependence of extreme values. 
 
  \begin{figure}[h]
	\begin{center}
		\includegraphics[height=4.5cm,width=12cm]{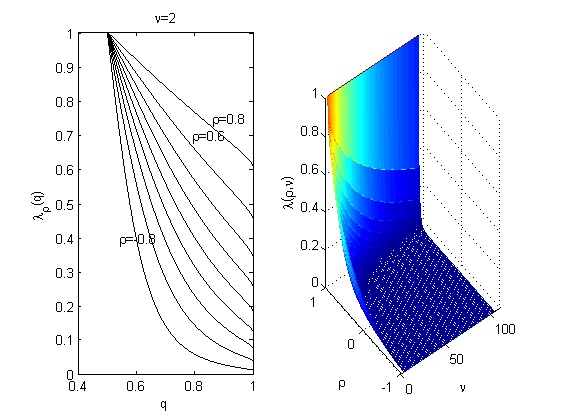}
	\end{center}
	\caption{ Tail Dependence Coefficient for t-Student Copula}
\end{figure}

Following  \citet{ronc2}(cf. e.g. \cite{ronc2} p. 299) , we can express the  $t$ copula $C^t_{\rho,\nu}\left(u,v\right)$ as follows:
\begin{align}
		C^t_{\rho,\nu}\left(u,v\right)= \int_0^u t_{\nu+1}\left(\left(\frac{\nu +1}{\nu+\left[t_\nu^{-1}\left(u\right)\right]^2}\right)^{1/2}   \frac{t_\nu^{-1}\left(v\right)-\rho t_\nu^{-1}\left(t\right)}{\sqrt{1-\left(\rho\right)^2}}\right)dt.
\end{align}
Now based on theorem~\ref{main} we compute the expression of $g\left(v, u\right)$. We obtain
\begin{align*}
	g\left(v, u\right) &=   \frac{\partial C^t_{\rho,\nu}\left(u,v\right)\left(u,v \right)}{\partial u }& \nonumber\\
	&=\frac{\partial\left(  \int_0^u{ t_{\nu+1}\left(\left(\frac{\nu +1}{\nu+\left[t_\nu^{-1}\left(u\right)\right]^2}\right)^{1/2}   \frac{t_\nu^{-1}\left(v\right)-\rho t_\nu^{-1}\left(t\right)}{\sqrt{1-\left(\rho\right)^2}}\right)}dt \right)}{\partial u }& \nonumber\\
&	=   t_{\nu+1}\left(\left(\frac{\nu +1}{\nu+\left[t_\nu^{-1}\left(u\right)\right]^2}\right)^{1/2}   \frac{t_\nu^{-1}\left(v\right)-\rho t_\nu^{-1}\left(u\right)}{\sqrt{1-\left(\rho\right)^2}}\right).&
\end{align*}
 The function $g$ is invertible and its inverse is obtained  by solving the equation
\begin{align*}
	g\left(v, u\right)= t_{\nu+1}\left(\left(\frac{\nu +1}{\nu+\left[t_\nu^{-1}\left(u\right)\right]^2}\right)^{1/2}   \frac{t_\nu^{-1}\left(v\right)-\rho t_\nu^{-1}\left(u\right)}{\sqrt{1-\left(\rho\right)^2}}\right)&=\alpha
	\end{align*}
for $v$.	This leads to,
	\begin{align*}
  v=g^{-1}\left(\alpha, u\right)=t_\nu\left( \rho t_\nu^{-1}\left(u\right)+\sqrt{\frac{\left(1-\left(\rho\right)^2\right)\left(\nu+\left[t_\nu^{-1}\left(u\right)\right]^2\right)}{\nu+1}}  t^{-1}_{\nu+1}\left(\alpha\right)  \right).
\end{align*}
We obtain the following formula for $CoVaR_{\alpha}^{s|L^i=l}$ and $CoVaR_{\alpha}^{\beta}$ when the dependence is modeling by a t-copula.
\begin{prop}

\begin{align*}
	CoVaR_\alpha^{s|L^i=l}%&=F^{-1}_s\left(g^{-1}\left(\alpha,F_i\left(l\right)\right)\right)& \nonumber\\
	                     & =F^{-1}_s\left(t_\nu\left( \rho_= t_\nu^{-1}\left(F_i\left(l\right)\right)+\sqrt{\frac{\left(1-\left(\rho_=\right)^2\right)\left(\nu+\left[t_\nu^{-1}\left(F_i\left(l\right)\right)\right]^2\right)}{\nu+1}}  t^{-1}_{\nu+1}\left(\alpha\right)  \right)\right) &
\end{align*}
 and
\begin{align*}
	CoVaR_\alpha^{\beta}=F^{-1}_s\left(t_\nu\left( \rho_= t_\nu^{-1}\left(\beta\right)+\sqrt{\frac{\left(1-\left(\rho_=\right)^2\right)\left(\nu+\left[t_\nu^{-1}\left(\beta\right)\right]^2\right)}{\nu+1}}  t^{-1}_{\nu+1}\left(\alpha\right)  \right)\right). &
\end{align*}
 
\end{prop}

Where $F^i$ and $F^s$ represent the univariate distribution function of $L^i$ and $L^s$ respectively. $\beta$ denotes the regulatory risk level of the financial institution $i$.\\

If $L^i$ and $L^s$ are t distributed with degrees of freedom $\nu$ then
\begin{align}
	CoVaR_\alpha^{s|L^i=l}%&=F^{-1}_s\left(g^{-1}\left(\alpha,F_i\left(l\right)\right)\right)& \nonumber\\
	                     & = \left( \rho_= \cdot l\right) +\sqrt{  \frac{\left(1-\left(\rho_=\right)^2\right)\left(\nu+ l^2\right)}{\nu+1}   }  t^{-1}_{\nu+1}\left(\alpha\right)   &
\end{align}

and we have in this case (cf. \cite{ext_risk} eq.6.B.31)

\begin{align*}
	\rho_= = \lim_{\Delta l \to 0} \frac{\rho}{\rho^2 + \frac{E\left[E\left({L^s}^2|L^i\right)-\rho^2 {L^i}^2 |L^i\in I\right]}{Var\left(L^i|L^i\in I\right)}} ,\ \ I = \left(l; l + \Delta l\right).
\end{align*}

Note that the  dependence in the Gaussian and t-copulas setting are essentially determined by the correlation coefficient $\rho$  (elliptical copula).
 The correlation coefficient  is often considered   as being a poor tool for describing dependence when the margins are  non-normal (cf. \cite{qrm}. This  motivates the next section.

%%%%%%%%%%%%%%%%%%%%%%%%%%%%%%%%%%%%%%%%%%%%%%%%%%%%%%
%%%%%%%%%%%%%%%%%%%%%%%%%%%%%%%%%%%%%%%%%%%%%%%%%%%%%

\subsection{Bivariate Archimedean Copulas}
\label{ex4}
We can give in this section a general expression of the  $CoVaR_{\alpha}^{s|L^i=l}$ for some  Archimedean Copulas.
Archimedean copulas are   often used in practice because of their analytical property, and ability to reproduce a large spectrum of dependence structures. Differently from the elliptical copulas, The definition of a bivariate copula are not derived from a given bivariate distribution.  The construction of Archimedean copulas is  based on special function (the so called \textsl{generator}). The generator of a Archimedean copula is a convex and strictly decreasing continuous function $\varphi$ from $\left[0 1\right]$ to $\left[0,\infty\right]$ with $\varphi\left(1\right)=0$.

\begin{mydef}[pseudo-inverse, cf. \cite{qrm} def.~5.41] 
define a pseudo-inverse of $\varphi$ with domain $\left[0,\infty\right]$ by

\begin{align*}
\varphi^{\left[-1\right]}\left(t\right)= \left\{ \begin{array}{rl}
  \varphi^{-1}\left(t\right) &\mbox{ if $0\leq t\leq \varphi\left(0\right)  $} \\
  0 &\mbox{ if $\varphi\left(0\right)< t \leq \infty $.}
       \end{array} \right.
\end{align*}
Note that the  composition of the  pseudo-inverse with the generator gives the identity i.e.
\begin{align*}
	\varphi^{\left[-1\right]}\left(\varphi\left(t\right)\right)=t.\   \  \forall \ t\in \left[0,\infty\right].
\end{align*}
\end{mydef} 
If $\varphi\left(0\right) = \infty$  the generator is said to be strict and it is equivalent to the ordinary functional inverse $\varphi^{-1}$.\\
Given a generator $\varphi$ we can construct the corresponding Archimedean copula as follows
\begin{align*}
	C(u, v) = \varphi^{\left[-1\right]}\left(\varphi(u)+\varphi(v)\right).
\end{align*}

The lower and upper tail dependence coefficient of an Archimedean copula  can be computed using following   corollary.

 \begin{co}[\cite{nelsen} co.~5.4.3] Let Let C be an Archimedean copula with  a   continuous, strictly, decreasing and  convex generator $\varphi$. Then
\begin{align*}
	 \lambda_u&= 2 - \lim_{x \to 0^+}\frac{1-\varphi^{-1}\left(2x\right)}{1- \varphi^{-1}\left(x\right)}&\\
 	\lambda_l&= \lim_{x \to \infty}\frac{1-\varphi^{-1}\left(2x\right)}{1- \varphi^{-1}\left(x\right)}&
\end{align*}
  \end{co}

\begin{prop}
Let C be an Archimedean copula with  a   continuous, strictly, decreasing and  convex  generator $\varphi$ i.e.
\begin{align*}
	C(u, v) = \varphi^{-1}\left(\varphi(u)+\varphi(v)\right).
\end{align*}
Then the function $g$  defined as in theorem \ref{main} is given by (cf. \cite{nelsen} Thm.~4.3.8):
\begin{align*}
	g\left(v,u\right)=\frac{\partial C\left(u, v \right)}{\partial u}= \frac{\varphi'\left(u\right)}{\varphi'\left(\varphi^{-1}\left[\varphi\left(u\right)+\varphi\left(v\right)\right]\right)}.
\end{align*}
\end{prop}

Set $g\left(v,u\right)=\alpha$ and solver for $v$, we obtain the inverse of $g$. Namely:
\begin{align*}
	g^{-1}\left(\alpha,u\right)= \varphi^{-1}\left( \varphi\left(\varphi'^{-1}\left(\frac{\varphi'\left(u\right)}{\alpha}\right)\right)-\varphi\left(u\right)\right).
\end{align*}

Based on theorem \ref{main} we derive the following proposition , which gives the  expression of $CoVaR_{\alpha}^{s|L^i=l}$ for some Archimedean copulas 

\begin{prop}
Let Let C be an Archimedean copula with  a   continuous, strictly, decreasing and  convex generator $\varphi$
Let $L^s$ and $L^i$ be two random variables representing the loss of the system $s$ and institution $i$  with joint distribution defined by a bivariate copula $C$ with  marginal distribution functions $F_s$ and $F_i$ respectively i.e.
\begin{align*}
	F_{L^i,L^s}\left(x,y\right)=C\left(F_i\left(x\right),F_s\left(y\right)\right).
\end{align*}
If C is an Archimedean copula with  a   continuous, strictly, decreasing and  convex generator $\varphi$, then
 the explicit (or closed) formula for the  $CoVaR_{\alpha}^{ s|L^i=l}$ at level $\alpha,\   0 < \alpha < 1 $ for a certain fixed value $l$ of $L^i$  is given by:
\begin{align*}
	CoVaR_{\alpha}^{ s|L^i=l}&=F_s^{-1}\left(g^{-1}\left(\alpha,F_i\left(l\right)\right)\right).\nonumber& \\
	                         &=F_s^{-1}\left(\varphi^{-1}\left( \varphi\left(\varphi'^{-1}\left(\frac{\varphi'\left(F_i\left(l\right)\right)}{\alpha}\right)\right)-\varphi\left(F_i\left(l\right)\right)\right)\right)&
\end{align*}
and
\begin{align*}
		CoVaR_{\alpha}^{\beta}=F_s^{-1}\left(\varphi^{-1}\left( \varphi\left(\varphi'^{-1}\left(\frac{\varphi'\left(          \beta \right)}{\alpha}\right)\right)-\varphi\left( \beta\right)\right)\right).
\end{align*}
\end{prop}
We are particular interested here  by the archimedean copulas showing   positive upper tail dependence(e.g. Gumbel copula).  
%%%%%%%%%%%%%%%%%%%%%%%%%%%%%%%%%%%%%%%%%%%%%%%%%%%%%%%%%%%%%%%%%%%%%%%%%%%%%%%%%%%%%%%%%%%%
\subsubsection{Gumbel copula}
\label{ex3}
\begin{mydef}
The bivariate Gumbel Copula function is given by (cf. \cite{nelsen} ex.~4.25)
\begin{align*}
	C_\theta^{Gu}\left(u,v\right)=exp\left(-\left[\left(-ln u\right)^{\theta}+ \left(-ln v\right)^{\theta}   \right]^{\frac{1}{\theta}}\right),      \ \ \  1\leq \theta < \infty,
\end{align*}
\end{mydef}
where $\theta $ represents the strength of dependence. Note that:
\begin{itemize}
	\item For $\theta = 1$ we have no dependency copula. i.e.  $	C_{\theta}^{Gu}\left(u,v\right)=uv$
	\item For $\theta \rightarrow \infty$ we have the perfect dependence i.e. 	$C_{\theta}^{Gu}\left(u,v\right)= min\left(u,v\right)=M\left(u,v\right)$ with m and M represented the Fréchet-Hoeffding lower and upper bound respectively.
\end{itemize}
The generator of the bivariate Gumbel  is given by $\varphi_{\theta}\left(t\right)=\left(-ln t\right)^{\theta} \ \ for \ \theta \geq 1$.
 %Such that we can represent the Gumbel copulas as:
%\begin{align*}
%	C_{\theta}\left(u, v\right)=\varphi^{-1}\left(\varphi\left(u\right)+\varphi\left(v\right)\right).
%\end{align*}

The tail dependence coefficient of  the Gumbel copula is therefore given by:  
\begin{align*}
	\lambda_u =2 - \lim_{x \to 0^+}\frac{1-\varphi^{-1}\left(2x\right)}{1- \varphi^{-1}\left(x\right)} =  2-2^{\frac{1}{\theta}},\  \   \   \   and \  \  \  \lambda_l = 0
\end{align*}
and we have
\begin{align}
	g\left(v,u\right)&=\frac{\partial C\left(u,v\right)}{\partial u} &\nonumber\\
	                 &=\frac{\partial  \  exp\left(-\left[\left(-ln u\right)^{\theta}+ \left(-ln v\right)^{\theta}   \right]^{\frac{1}{\theta}}\right)}{\partial u}& \nonumber \\
	                 &= exp\left(-\left[\left(-ln u\right)^{\theta}+ \left(-ln v\right)^{\theta}  \right]^{\frac{1}{\theta}}\right) \cdot &  \nonumber \\
	                 &\left(\left(-ln u\right)^{\theta}+ \left(-ln v\right)^{\theta}   \right)\cdot  & \nonumber\\
	                 &  \frac{\left(-ln u\right)^{\theta-1}}{u} .  &   \label{eq:54} 
\end{align}
Note that \eqref{eq:54} is a strictly increasing with respect to $v$. Its   inverse  $g^{-1}\left(\alpha, u \right)$ is thus well defined. However its inverse  $g^{-1}\left(\alpha, u \right)$ cannot be expressed in an explicit form. Hence we cannot derive $CoVaR_{\alpha}^{s|L^i=l}$ analytically, but we can use in this case we can use numerical methods.

%%%%%%%%%%%%%%%%%%%%%%%%%%%%%%%%%%%%%%%%%%%%%%%%%%%%%%%%%%
\subsubsection{Clayton Copula}
The generator of the bivariate Clayton copula is given by:
\begin{align*}
	\varphi\left(t\right)= t^{-\theta}-1.
\end{align*}
According to proposition 6 we have the following expression for the $CoVaR_{\alpha}^{ s|L^i=l}$
\begin{align*}
	CoVaR_{\alpha}^{ s|L^i=l}  &=F_s^{-1}\left( \left[\left(\alpha^{-\frac{\theta}{1+\theta}}-1\right)\left(F_i\left(l\right)\right)^{-\theta}+1\right]^{-\frac{1}{\theta}}\right).
\end{align*}

%%%%%%%%%%%%%%%%%%%%%%%%%%%%%%%%%%%%%%%%%%%%%%%%%%%%%%%%%%%%%%%%%

%%%%%%%%%%%%%%%%%%%%%%%%%%%%%%%%%%%%%%%%%%%%%%%%%%%%%%%%%%%%%%%%%

\section{Conclusion}
Managing and regulating the systemic risk is a fundamental problem for financial regulators and risk managers especially in the context of the current crisis. 
The must important challenge here is the modeling and the quantification of the potential contribution of one given individual financial institution to the financial system. One of the main approaches  to solve this problem is the $covar$ method proposed by Adrian and Brunnermeier in \cite{covar}. Where the financial system is defined as a portfolio of two items such that  the loss of the system is represented by a random vector $\left(L^i \ L^s\right)$ where $L^i$ is the loss of the focused financial institution $i$ and $L^s$ the loss of the financial system $s$, and  the marginal risk contribution of the bank $i$  to systemic risk $s$ is quantify  by the risk measure $\Delta CoVaR^{s|i}$ which is defined as the  difference between 	$CoVaR_{\alpha}^{ s|L^i=VaR_i}$ and $CoVaR_{\alpha}^{ s|L^i=E\left(L^i\right)}$ (see def.~\ref{d4} ). The problem of the computation and the analysis of the term $CoVaR_{\alpha}^{ s|L^i=l}$ for a given $l$   is thus very  important for the implementation of $CoVaR$ especially in the non-Gaussian world, but still we do not get any definite solution.   
As an answer to this problem, we have developed our method, based on copula theory, an analytical framework for  the implementation of the $CoVaR$ methodology   where  the risk measure $CoVaR_{\alpha}^{ s|L^i=l}$ is expressed in a closed form in terms of  the marginal distributions  $F_s$ and  $F_i$ separately and  the copula $C$ between the focused  financial institution $i$ and the financial system $s$. This framework provides an effective computation and  analysis tool for the systemic risk using $CoVaR$ for a widely used class of distribution function and comovement dynamic(see Theorem~\ref{main}), which captures not only linear correlation but also nonlinear tail dependencies between the banks in one financial system  (which summarise the main features of loss distribution) as opposed to the "linear quantile regression" and the formula in \cite{manfred}  where the dependence is modeled only by the linear correlation coefficient. In fact our approach allows to analyse  the marginal effect of $F_i$, $F_s$ and $C$ of the systemic risk. We show for example the systemic risk contribution of $i$ is independent  of $F_i$ (see \eqref{eq:23}) and  highlight  in remark~\ref{r6} some properties of $CoVaR$ according to the nature of $F_s$.  % We also give the condition that a copula have to satisfy by the analysis or computing of systemic risk contribution (see Section~\ref{sec:4}).  
Our approach  can also be used to develop  closed formulas for  the computation of related macro-risk measures like $CoVaR^{i|s}$(cf. \cite{covar}), $\Delta CoVaR^{s|i}$, $\Delta CollVaR^{s|i}$ (cf. \cite{manfred}), and $\Delta CondVaR^{s|i}$ (cf. \cite{manfred}).

%By taking this in consideration we have:
%\begin{align*}
	%f_{y|x}\left(y\right)=\left( \frac{1-\rho^2}{\left(\frac{\nu +1}{\nu + x^2}\right)}\right)^{\frac{1}{2}} t_{\nu+1}\left(r\right). 
%\end{align*}

 %This implies that

%\begin{align*}
 %t_{\nu+1}\left(r\right)&= 	f_{y|x}\left(y\right)	\left( \frac{1-\rho^2}{\left(\frac{\nu +1}{\nu + x^2}\right)}\right)^{-\frac{1}{2}}&. \\
  %                      &=\frac{\Gamma\left(\frac{\nu +2}{2}\right)}{\sqrt{\pi\left(\nu+1\right)}\Gamma\left(\frac{\nu+1}{2}\right)} \left(1+ \frac{r^2}{\nu +1}\right)^{-\frac{\nu+2}{2}}&
%\end{align*}

 \nocite{Bar2012 }
 \nocite{Demarta05thet}
 \nocite{dirk}
 %\nocite{taraschev}
% \nocite{Y_analyticalcovar}
 \nocite{ash}
 \nocite{cbidis}
 \nocite{carol}
 \nocite{dinunno}
 \nocite{hull2}
 \nocite{er}
 \nocite{multistat}
 \nocite{qfrm}
 \nocite{qfrm_s}
 \nocite{darsow}
 \nocite{ec}
 \nocite{coherent}
 \nocite{gordy}
 \nocite{lut}
 \nocite{htd}
 \nocite{engel1}
 \nocite{copfin}
 \nocite{pat}
 \nocite{guenther}
 \nocite{multi_t}
 \nocite{emb}
 \nocite{hong}
 \nocite{excop}
 \nocite{lehar1}
 \nocite{ronc2}
 \nocite{Brahimi}
 \nocite{Albrecht}
 \nocite{acharya}
 \nocite{Brahimi2}
 \nocite{engel2}
 \nocite{shiryaev}
 \nocite{roh}
 \nocite{Embrechts99correlationand}
  \nocite{brian}

\textbf{Reference}\\

%\bibliographystyle{model1a-num-names}


\bibliographystyle{abbrvnat}
\bibliography{ryan}

\end{document}